\documentclass{article}
\usepackage{hiph-preprint}
\usepackage{graphicx}
\volnumber{22} \issuenumber{1} \edyear{2005}                             
\frompage{000} \topage{000}                                              
\def\rightmark{Chiral restoration}
\def\leftmark{T. Tatsumi and E. Nakano}

\newcommand\beq{\begin{equation}}
\newcommand\eeq{\end{equation}}
\newcommand\beqa{\begin{eqnarray}}
\newcommand\eeqa{\end{eqnarray}}

\newcommand{\ds}[1]{#1 \hspace{-0.5em}/}  

\newcommand\bgamma{\mbox{\boldmath$\gamma$}}

\title{How is chiral symmetry restored at finite density?} 
\authors{ 
{Toshitaka Tatsumi$^1$ and Eiji Nakano$^{2,a}$ %
\index{Tatsumi, T.} 
\index{Nakano, E.} 
}\\[2.812mm]
{\normalsize
\hspace*{-8pt}$^1$ Department of Physics, Kyoto University,  
Kyoto 606-8502, Japan\\[0.2ex] 
\hspace*{-8pt}$^2$ Yukawa Institute for Theoretical Physics, 
Kyoto University, Kyoto 606-8502, Japan
}}
 
\abstract{Taking into account pseudoscalar as well as scalar condensates,
we reexamine 
the chiral restoration path on the chiral manifold. We shall see both
condensates 
coherently produce a density wave at a certain density, which delays  
 chiral
restoration as density or temperature is increased.}

 
\makeindex
\begin{document}

\maketitle

\section{Introduction}\label{intro}
 Condensed matter physics of QCD is now one of the active fields in
 nuclear physics and the main purpose is to 
figure out the phase diagram in the temperature-density plane. 
Color superconductivity
is realized by forming the condensation of
 the quark 
Cooper pairs. There should be also the instability by creating particle-hole
 ({\it p-h}) pairs near the 
Fermi surface. 
This is a condensation of 
 {\it p-h} pairs, where the standing wave with the wave number $q$ is
 developed in the ground state. 
The condensate with $q=0$ is uniform, and, e.g.,  ferromagnetic instability
 corresponds 
to this case\cite{tat}. A typical example of the condensate with $q=2k_F$
 ($k_F$: Fermi momentum) is 
the Overhauser state or the spin-density wave state\cite{ove}. 
In quark matter the scalar or tensor density may form a standing wave
 with 
a finite momentum about $2k_F$, called the chiral density wave\cite{chi}. 
 We consider here another type of the density wave due to chiral
 symmetry in QCD \cite{nak}.
 
\section{Dual Chiral Density Wave }\label{dcdw}  
We are concentrated 
on the flavor-$SU(2)$ case, and introduce  
a {\it dual chiral-density wave} (DCDW) state, where both the scalar and
pseudo-scalar densities always 
reside on the hypersphere with a constant modulus $\Delta$, 
$\langle\bar\psi
\psi\rangle^2+\langle\bar\psi i\gamma_5\tau_i\psi\rangle^2=\Delta^2$,  
while each density is {\it spatially non-uniform};  
we consider the following configuration in quark matter,
\begin{eqnarray}
\langle\bar\psi\psi\rangle &=& \Delta\cos\theta({\bf r})\nonumber\\
\langle\bar\psi i\gamma_5\tau_3\psi\rangle&=&\Delta\sin\theta({\bf r}).
\label{gene1}
\end{eqnarray}
Taking the simplest but  nontrivial form for the 
chiral angle $\theta$ such that $\theta({\bf r})={\bf q\cdot r}$,
we call this configuration DCDW.  

It should be noted that 
we can construct the DCDW state by acting a space-dependent 
chiral transformation such as 
$
\psi \longrightarrow \psi_W=\exp(i\gamma_5\tau_3\theta({\bf r})/2)\psi,
$
on the usual spontaneously symmetry breaking (SSB) state 
where only the scalar density condenses. 
When the chiral angle has some spatial dependence, 
there should appear an extra term in the Lagrangian 
by the local chiral transformation,  
$\bar{\psi}_W (\gamma_5\tau_3\bgamma\cdot\nabla\theta/2) \psi_W$, 
due to the non-commutability of $\theta({\bf r})$ with the kinetic
(differential) operator. 
Consequently some terms are added in the effective action by this contribution.
One is the interaction term between quarks inside the Fermi sea and DCDW.
The other is rather nontrivial and comes from the 
vacuum-polarization effect: it should give an 
additional term, $\propto (\nabla\theta)^2$ in the lowest order.
This can be regarded as an appearance of the kinetic energy term for
DCDW through the vacuum polarization \cite{sug}.  
Therefore,  
when the interaction energy is superior to the kinetic energy, quark
matter becomes unstable to form DCDW.


We will see that 
the mechanism is quite similar to that for the spin density wave 
suggested by Overhauser \cite{ove}, and entirely reflects 
the nesting of the Fermi surface . 
In the higher dimensional systems, however, the nesting is incomplete
and the density wave should be
formed 
provided the interaction of a relevant {\it p-h} channel is strong enough.

\section{Phase diagram within the NJL model}\label{details}
We explicitly demonstrate that quark matter 
becomes unstable for a formation of DCDW at moderate densities, 
using the Nambu-Jona-Lasinio (NJL) model. 

Assuming here the form (\ref{gene1}) for the mean-fields, 
we define a new quark field $\psi_W$ by the Weinberg 
transformation.
In terms of the new field the NJL Lagrangian 
renders
\beq
{\cal L}_{MF}=\bar\psi_W[i\ds{\partial}-M-1/2\gamma_5\tau_3\ds{q}]\psi_W
-G\Delta^2,
\label{effl}
\eeq
which appears to be the same as the usual one 
except an ``external'' 
axial-vector field, ${\bf q}$; the {\it amplitude} of DCDW 
generates the dynamical quark mass $M=-2G\Delta$ in this case. 
The Dirac equation for $\psi_W$ then gives a 
{\it spatially uniform} solution  
\footnote{This feature is very different from refs.\cite{chi}, where the wave 
function is no more plane wave.}
with the eigenvalues 
\beq
E^{\pm}({\bf k})=\sqrt{E_{k}^{2}+|{\bf q}|^2/4\pm \sqrt{({\bf
k}\cdot{\bf q})^2+M^{2}|{\bf q}|^2}},~~~E_k=(M^2+|{\bf k}|^2)^{1/2}
\label{energy}
\eeq
for positive energy (valence) quarks with different polarizations
denoted by the sign $\pm$. For 
negative energy quarks in the Dirac sea, they have an energy spectrum 
symmetric with 
respect to null line because of charge conjugation symmetry in the  
Lagrangian (\ref{effl}). The energy spectra (\ref{energy}) show a
salient feature: they break  
rotation symmetry, and lead to the axial-symmetrically deformed Fermi seas.

Hereafter, we choose ${\bf q}//\hat z$, ${\bf q}=(0,0,q)$ with $q>0$, 
without loss of generality. 
The effective potential is then given by summing up all the energy levels, 
\beqa
\Omega_{\rm total}=\Omega_{\rm val}+\Omega_{\rm vac}+M^2/4G, 
\label{therm}
\eeqa
The first term 
$\Omega_{\rm val}$ is the 
contribution by the valence quarks filled up to the Fermi energy 
in each Fermi sea, 
while the second term $\Omega_{\rm vac}$ is the vacuum 
contribution that is formally divergent and should be properly
regularized. 
Note that we cannot apply the usual momentum cut-off 
regularization scheme to $\Omega_{\rm{vac}}$, 
since the energy spectrum has no more rotation symmetry. 
Instead, we adopt the proper-time regularization scheme, 
which may be a most suitable one for our
purpose\cite{nak}. Then $\Omega_{\rm{vac}}$ results in a function of $q$
and its lowest order contribution is proportional to $q^2$.

We can expect that the optimal value of $q$ is $2k_F$ in 1+1 dimensions. 
First, consider the energy spectra with ${\bf k}=(0,0,k_z)$ for massless quarks, $M=0$. 
Two energy spectra are essentially reduced to the usual ones 
$E^\pm({\bf k})=|{\bf k}|$ in this case.
Then we can see
a level crossing at  $k_z=0$. Once the mass $M$ is taken into 
account, this degeneracy is resolved and the energy gap appears there. 
Hence  we have always an energy gain by filling only the
lower-energy spectrum $E^-({\bf k})$ up to the Fermi energy, if the relation 
,$q=2k_F$, holds. However, we shall see that $q<2k_F$ in 3+1 dimensions by
numerical evaluations.
This  
mechanism is very similar to that of spin density wave 
by Overhauser \cite{ove}
\footnote{The similar mechanism 
also works in the context of the chiral density 
wave \cite{chi}.}
.

\begin{figure}[h]
\begin{center}
\includegraphics[height=4cm]{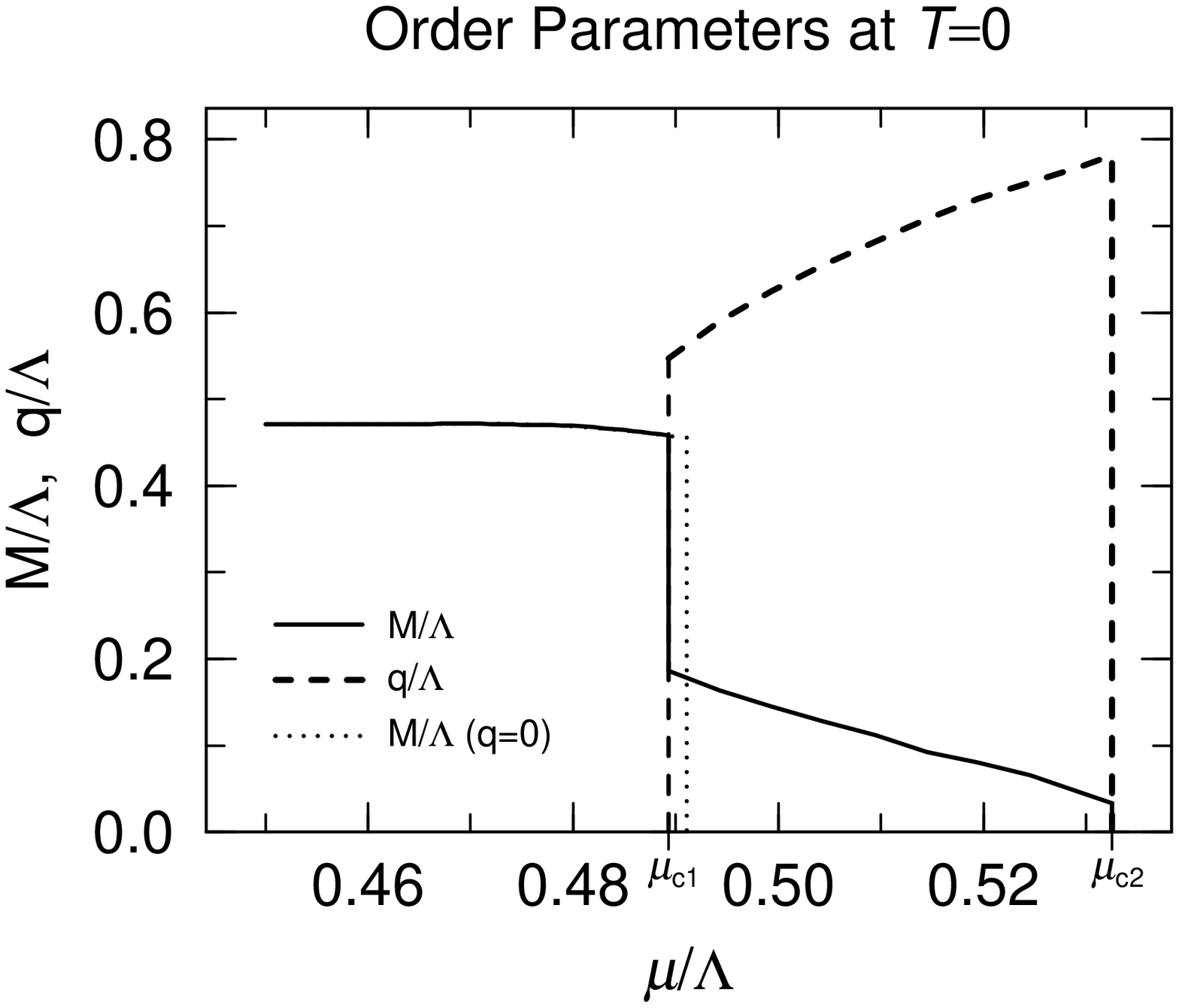}
\includegraphics[height=4cm]{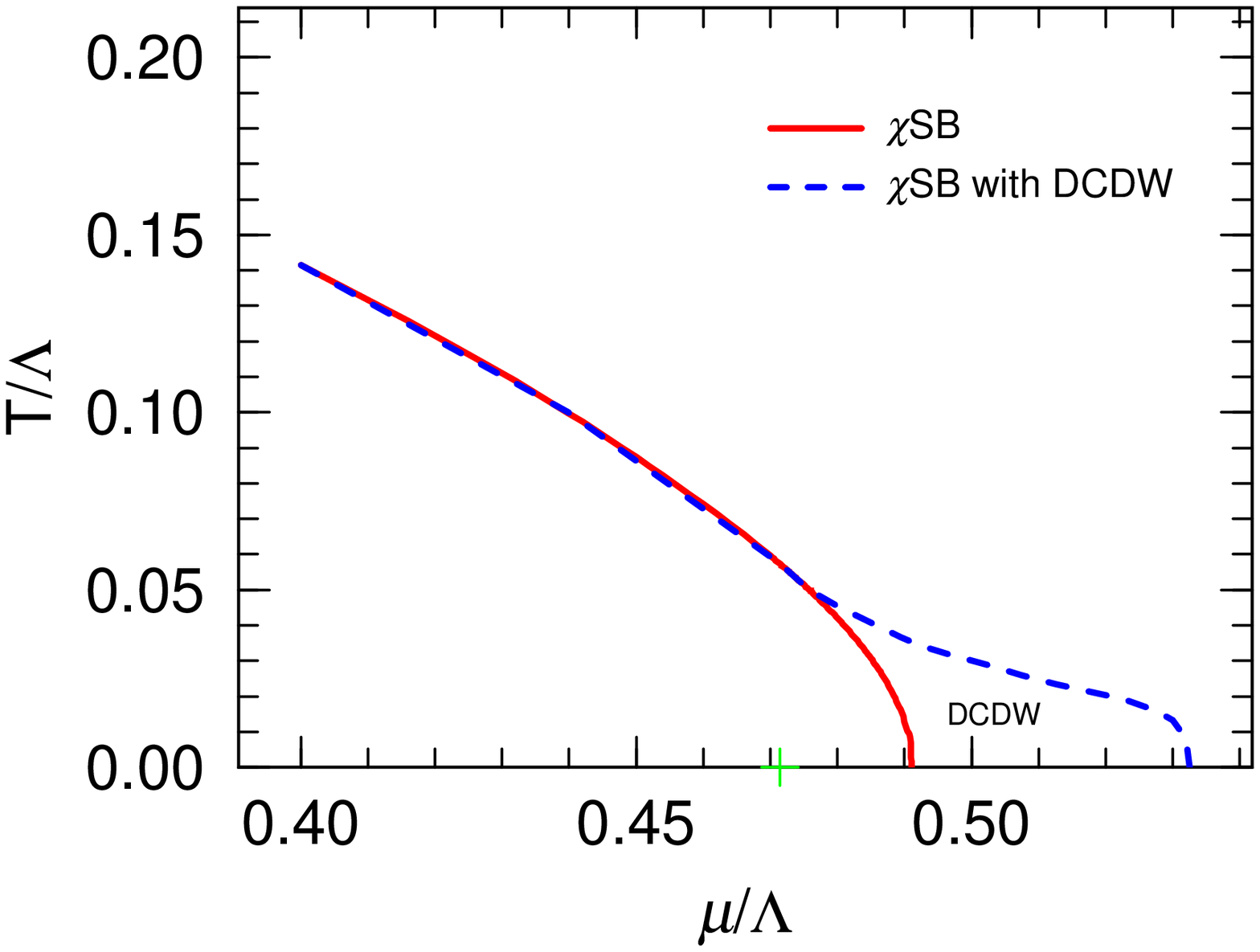}
\end{center}
\vspace*{-0.8cm}
\caption{Left panel: the wave number $q$ and the dynamical mass $M$ are plotted 
as functions of $\mu$. 
Solid (dotted) line for $M$ with (without) DCDW, 
and dashed line for $q$. Right panel: the phase diagram in the
 temperature-density plane.}
\label{op1}
\vspace*{-0.1cm}
\end{figure}

The left panel of Fig.~\ref{op1} demonstrates the behaviors of the order-parameters $M$ and $q$ 
as functions of chemical potential $\mu$ at zero temperature 
for a parameter set $G \Lambda^2=6, \Lambda=860$ MeV. 
%
It is found that the magnitude of $q$ becomes finite 
just before the ordinary chiral-symmetry restoration, 
and DCDW survives in the finite range of $\mu$,  
which corresponds to the baryon-number densities $\rho_b/\rho_0=3.62-5.30$. 
The wave number $q$ increases with density, 
while its value is always smaller than the canonical value of $2k_F$  
due to the higher dimensional effects. 

The right panel shows a phase diagram in the temperature-density
plane. We can see that DCDW develops outside the phase boundary for the
usual SSB phase. 
We thus conclude that DCDW is induced by finite-density contributions, 
and has the effect to extend the SSB phase ($M\neq 0$) to high density
region, which suggests another path for chiral-symmetry restoration by
way of the DCDW state at finite density.

Finally we would like to indicate an interesting magnetic aspect of the 
DCDW state. 
The magnetic moment is evaluated to be spatially oscillating like the 
{\it spin density wave} \cite{nak}. 

\section{Summary and concluding remarks}\label{maths}

We have seen that quark matter becomes soft for producing DCDW at a
certain density just near the usual chiral restoration, which gives
another path for the chiral restoration. 
This is due to the nesting effect of the Fermi sea. 

It would be interesting to recall that DCDW  
is similar to pion condensation \cite{dau}, 
where 
meson condensates take the same form as Eq.~(\ref{gene1}). So we might
be tempted to connect pion condensation with DCDW by a symmetry
consideration (hadron-quark continuity). 

If DCDW phase is developed, we can expect a Nambu-Goldstone mode as {\it
phason}. Such phasons may affect the thermodynamic properties of quark matter.

This work is supported by the Japanese 
Grant-in-Aid for Scientific
Research Fund of the Ministry of Education, Culture, Sports, Science and
Technology (13640282, 16540246).

\vfill\eject
\end{document}